\documentclass[sigconf]{acmart}

\renewcommand\footnotetextcopyrightpermission[1]{} % removes footnote with conference information in first column
\usepackage[T1]{fontenc}
\usepackage[utf8]{inputenc}

\usepackage{amsmath}
\usepackage{lipsum}
\usepackage{graphicx}
\usepackage{caption}
\usepackage{hyperref}
\usepackage[skip=0.5ex]{subcaption}
\usepackage[linesnumbered, ruled]{algorithm2e}

\begin{document}

\title{Emergent Crowd Grouping via Heuristic Self-Organization}

\author{Xiao-Cheng Liao$^\dagger$, Wei-Neng Chen$^{\S*}$, Xiang-Ling Chen$^\S$ and Yi Mei$^\dagger$}
\thanks{*Wei-Neng Chen is the corresponding author.}
\affiliation{%
  \institution{}
  \department{$^\dagger$Centre for Data Science and Artificial Intelligence \& School of Engineering and Computer Science, \\Victoria University of Wellington, Wellington, New Zealand}
  \department{$^\S$School of Computer Science and Engineering, South China University of Technology, Guangzhou, China}
  \city{}
  \country{}
  $^\dagger$\{xiaocheng, yi.mei\}@ecs.vuw.ac.nz, $^\S$cschenwn@scut.edu.cn, $^\S$linger.xlchen@gmail.com
}
% \email{}

\renewcommand{\shortauthors}{Liao, et al.}

\begin{abstract}
Modeling crowds has many important applications in games and computer animation.
Inspired by the emergent following effect in real-life crowd scenarios, in this work, we develop a method for implicitly grouping moving agents.
We achieve this by analyzing local information around each agent and rotating its preferred velocity accordingly.
Each agent could automatically form an implicit group with its neighboring agents that have similar directions.
In contrast to an explicit group, there are no strict boundaries for an implicit group. 
If an agent's direction deviates from its group as a result of positional changes, it will autonomously exit the group or join another implicitly formed neighboring group.
This implicit grouping is autonomously emergent among agents rather than deliberately controlled by the algorithm.
The proposed method is compared with many crowd simulation models, and the experimental results indicate that our approach achieves the lowest congestion levels in some classic scenarios.
In addition, we demonstrate that adjusting the preferred velocity of agents can actually reduce the dissimilarity between their actual velocity and the original preferred velocity.
Our work is available online\footnote{https://pypi.org/project/pyfollower/}.
\end{abstract}

%%
%% Generate your CCSCML using http://dl.acm.org/ccs.cfm.
%%



\begin{teaserfigure}
  \includegraphics[width=\textwidth]{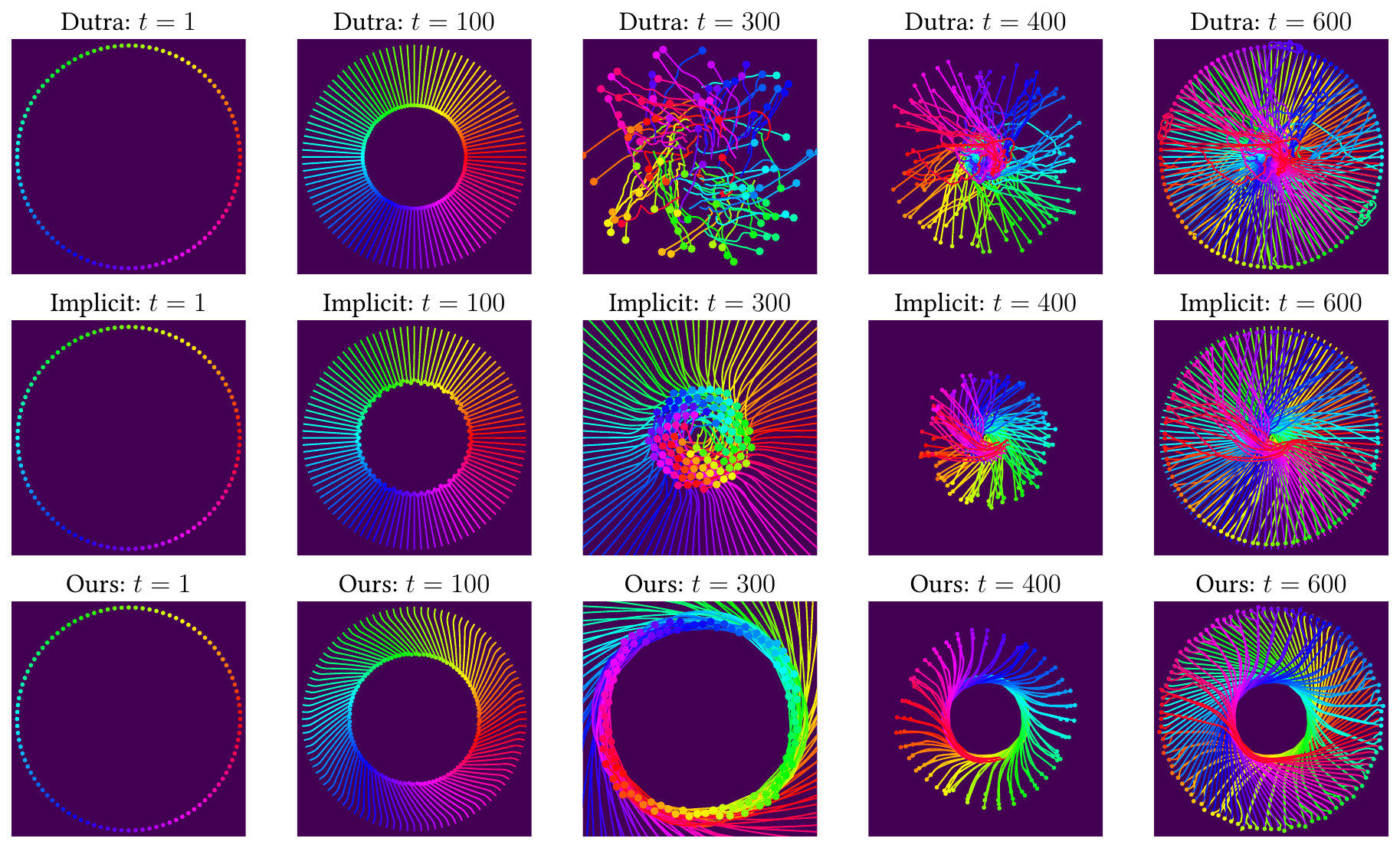}
  \caption{A classic scenario where 100 agents are distributed evenly on a circle, and their destination is to navigate to the antipodal position on the circle. In contrast to the severe trajectory oscillations produced by the Dutra method \cite{dutra2017gradient} at agent meeting area and the congestion with slow movement generated by the implicit crowd \cite{karamouzas2017implicit} in the same locations, our method achieved organized motion, realizing congestion-free scenarios. 
  Due to the symmetry in the motion of all agents, they self-organize into the same group, moving in the ``same direction'' (clockwise rotation).}
  \label{fig:teaser}
\end{teaserfigure}

\maketitle

\section{Introduction}

Large multi-agent systems, such as the real-time simulation of human crowds, are important tasks in games \cite{colas2022interaction, wolinski2014parameter}, animation \cite{gustafson2016mure, zhang2023orcanet} and visual effects \cite{kanyuk2015headstrong, van2021algorithms, hu2021heterogeneous}.
In the realm of crowd simulation studies, the classification typically involves two categories: flow-based, wherein the entire crowd is conceptualized as a substance with fluid-like characteristics, and agent-based, where each individual of the crowd is represented as an intelligent agent.
The focus of this work is specifically on microscopic approaches, delving into the intricacies of addressing the crowd simulation problem at the individual scale.
This work focuses on agent-based approaches that generate crowd dynamics from the micro perspective of individuals.

In agent-based methods, the primary focus is to model local interactions among agents which specifies how the movement of each agent is impacted by the presence of neighboring agents and obstacles.
Subsequently, macroscopic collective behaviors emerge through the microscopic interactions among individuals.
Amidst various modeling considerations, a significant portion \cite{van2020generalized,van2011reciprocal,lee2018crowd} is dedicated to addressing collision avoidance.
Typically, a local navigation algorithm needs to calculate a velocity for an agent at a given moment, enabling it to navigate its path safely around other nearby agents and obstacles without collisions.
In many modern techniques for procedural animation of intelligent systems and crowd dynamics, this local navigation algorithm could be optimizing a cost or energy function \cite{karamouzas2017implicit}, performing low-dimensional linear programming under the constraint of velocity obstacles \cite{fiorini1998motion, van2008reciprocal} and leaning to interact with agents\cite{charalambous2023greil,talukdar2023learning,lee2018crowd,kwiatkowski2023reward} 

In addition to the necessary collision avoidance characteristics, human crowds usually display a rich variety of self-organized behaviors that support efficient motion under everyday conditions \cite{moussaid2011simple}. One of the best-known examples is the spontaneous formation of uni-directional lanes in bi-directional pedestrian flows \cite{helbing1995social}. This intricate ability of crowds to self-organize has not only practical implications for everyday scenarios but also holds promise for optimizing crowd flow in controlled environments.

Group dynamics modeling is an effective crowd organizing method that reduces internal conflicts among crowds by organizing individuals into groups \cite{gorrini2014empirical, li2017grouping,nicolas2023social}.
Existing crowd simulation methods for modeling group behaviors often require explicit grouping or the establishment of heterogeneous leaders. 
Previous research on modeling group behavior has predominantly focused on cohesive motion or spatial structures within the crowd. 
Fundamental algorithms involve clustering agents into predefined groups, with each group maintaining a constant size (static grouping) \cite{bruneau2015going, tang2023crowd}. 
However, these methods fall short in capturing the evolving sizes of groups, dividing large crowds into subgroups, or merging smaller groups into larger ones \cite{he2016dynamic}.
Therefore, how to produce group dynamics via self-organization need further exploration.

In response to this limitation, this work proposes a method to facilitate the spontaneous emergence of group dynamics in crowds. 
The primary objective of this paper is to facilitate the emergence of self-organizing phenomena by adjusting the preferred velocity of agents, thereby improving the efficiency of crowd movement.
In terms of collision avoidance, Optimal Reciprocal Collision Avoidance (ORCA) \cite{van2011reciprocal} is employed as the underlying model, where the preferred velocity is a crucial variable compared to other collision avoidance models.
Our approach enables agents with similar forward directions to implicitly organize into groups. 
As agents move, changes in their positions result in individuals diverging automatically from the group when their target directions deviate. 
The formation, merging, splitting, and redistribution of these implicit groups are spontaneously driven by interactions among agents.
This paper achieves the emergence of automatic implicit grouping behavior in crowds through a bottom-up approach, potentially offering a novel perspective in the realm of crowd simulation.
The main contributions of this paper are as follows:

\begin{enumerate}
    \item Focusing on promoting the emergence of group dynamics, this paper proposes a method that allows agents to make slight adjustments to their preferred velocity in the direction based on the surrounding information. This enables the crowd to autonomously generate implicit groups, which can reduce internal conflicts among intelligent agents experiencing significant disparities in movement directions.
    
    \item The proposed method is compared with many agent-based crowd models, including the state-of-the-art method, implicit crowd \cite{karamouzas2017implicit}, 
    a crowd simulation model that utilizes implicit integration to produce highly smooth trajectories while maintaining strong robustness.
    Experimental results indicate that the proposed method performs better than compared algorithms in terms of crowd congestion level and the alignment between preferred velocity and actual velocity.
   
    \item We encapsulated our method into a Python package, accessible via the Pip package management tool. This facilitates easy integration and usage for researchers seeking to apply the developed approach in their work.

\end{enumerate}

\section{Background}
\subsection{Related Work}

Social Force model \cite{helbing1995social} is one of the earliest proposed models for simulating crowd dynamics.
Each agent in this model is influenced by surrounding agents and obstacles. This influence is defined as virtual physical forces, and, therefore, the movement of agents can be calculated based on principles of mechanics.
The concept of Velocity Obstacle (VO) space, denoting the velocities of an agent that would result in a collision between the agent and an obstacle or another agent at some future time, was introduced in \cite{fiorini1998motion}.
Therefore, collision avoidance can be transformed into a linear programming problem, considering velocity obstacles with respect to other agents or obstacles.
Numerous subsequent studies, building upon the concept of VO, have been proposed.
Rather than opting for a new velocity outside the velocity obstacle of the other agent in VO method, Reciprocal Velocity Obstacle (RVO) \cite{van2008reciprocal} selects a different velocity by averaging its current velocity with a velocity positioned outside the velocity obstacle of the other agent.
Compared to the original VO method, RVO can generate safer and smoother trajectories.
The formulation of RVO can only guarantee collision-free in certain specific conditions and Optimal Reciprocal Collision Avoidance (ORCA) was proposed to provide a sufficient condition for multiple agents to avoid collision \cite{van2011reciprocal}.
Some studies \cite{karamouzas2009predictive, karamouzas2014universal} also use the notion of time to collision to achieve collision avoidance and describe human interactions.
Vision information was also useful in agent-based crowd simulation and was explored in many studies \cite{ondvrej2010synthetic, hughes2015davis, dutra2017gradient}.
In these methods, each agent movement can be calculated by evaluating the future collisions based on the time to collision, time to closet approach or distance of closest approach of each pixel.
Karamouzas et al. \cite{karamouzas2017implicit} presented an optimization integrator named "Implicit Crowds," tailored to harmonize with nonlinear, anticipatory forces, and physics-based animation in multi-agent systems. Through the incorporation of dissipation functions, this approach demonstrates insensitivity to parameters and achieves the generation of remarkably smooth and collision-free trajectories for crowd movement.
In recent years, with the continuous increase in computational power, some researchers tried to use machine learning \cite{alahi2016social, zhong2022data,hu2021heterogeneous,kwiatkowski2023reward} to discover crowd behavior patterns and employ reinforcement learning \cite{lee2018crowd, talukdar2023learning, charalambous2023greil} to explore interaction strategies within crowds.
The above algorithms effectively model crowds and are sufficient to reproduce many known crowd phenomena. 
However, limited attention has been given to how the emergence of self-organizing phenomena in crowds can be facilitated through the local interactions of agents.

There is also some research on group dynamics.
Bruneau et al.\cite{bruneau2015going} studied collision avoidance among predefined and fixed pedestrian groups.
Tang et al. \cite{tang2023crowd} group individuals via machine learning methods.
Dynamic grouping \cite{he2016dynamic} was also explored to introduce a process for maintaining and updating groupings within the crowd after agents update their positions and velocities.
Despite these efforts, many real-world scenarios involve crowds exhibiting spontaneous cooperative behaviors, such as the emergence of channeling effects, without explicit grouping and criteria for grouping.
Further investigation is required into the emergence of crowd grouping.

\subsection{Agent-based Crowd Simulation}
\newcommand{\V}[1]{\mathbf{#1}}

In the setting of agent-based crowd simulation, the task involves navigating $N$ agents to their individual destinations while avoiding collisions with other agents or obstacles. 
These agents traverse a two dimensional plane and are commonly depicted as discs with radii $r$. 
The simulation includes a loop with frames of a fixed duration $\Delta t$.
At the simulation time step $t$, the state of agents can be characterized by three $1 \times N$ matrixes:
\begin{gather}
\V{X}^t = \left[ \V{x}_1^t, \V{x}_2^t, \cdots, \V{x}_i^t, \cdots, \V{x}_N^t \right] \\
\V{V}^t = \left[ \V{v}_1^t, \V{v}_2^t, \cdots, \V{v}_i^t, \cdots, \V{v}_N^t \right] \\
\V{P}^t = \left[ \V{p}_1^t, \V{p}_2^t, \cdots, \V{p}_i^t, \cdots, \V{p}_N^t \right] 
\end{gather}
where each $\V{x}_i^t$,$ \V{v}_i^t$,$ \V{p}_i^t$ represent the position, velocity, and preferred velocity of agent $A_i$ at time step $t$, respectively.

The preferred velocity signifies the direction in which the agent intends to move forward.
In its most basic form, the preferred velocity for each agent is represented as a vector starting from the agent's current position and pointing towards a local destination. 
Its magnitude is equal to a user-specified preferred speed $S_{pref}$.

Typically, a global path planner is assumed to operate outside local navigation, calculating a global path for each agent in the scenario.
This path can be decomposed into multiple local destinations.
Then, at each simulation time step $t$, the preferred velocities $\V{V}^t$ for all agents can be determined based on their current positions and local destinations.

To ascertain the state of agents in the subsequent time step $t+1$, local navigation is required to calculate $\V{X}^{t+1}$ and $\V{V}^{t+1}$ for the agents first. 
Then, $\V{P}^{t+1}$ can be determined based on current positions $\V{X}^t$ and the local destinations of agents.
Usually, $\V{X}^{t+1}$ can be calculated from $\V{V}^{t+1}$, and $\V{V}^{t+1}$ also can be inferred from $\V{X}^{t+1}$.
Thus, local navigation only needs to compute either of these two.
For example, in \cite{karamouzas2017implicit}, the calculation is for $\V{X}^{t+1}$, while in \cite{van2011reciprocal,dutra2017gradient,van2008reciprocal}, the calculation is for $\V{V}^{t+1}$.

When the surrounding environment is crowded, 
the actual velocity $\V{V}^{t+1}$ of agents often deviates from their preferred velocity $\V{V}^t$.
Agents aspire to advance towards their local goals but are required to adhere to certain local navigation constraints, such as avoiding collisions with other agents, maintaining proximity to a specific agent or group, and other relevant considerations.
According to some crowd simulation models, such as \cite{van2008reciprocal,karamouzas2014universal,karamouzas2017implicit}, the agent repeatedly makes a local adjustment to step in a certain direction.
In this work, for convenience, we use a function to represent this local adjustment/navigation:
\begin{equation}
\V{V}^{t+1} = \mathcal{F}\left( \V{X}^{t}, \V{V}^{t}, \V{P}^t \right).
\end{equation}
where $\mathcal{F}(\cdot)$ serves as an abstraction for the process of local navigation algorithms (collision avoidance in particular).
It take the state of agents as input and output the velocities of all agents of the next time step.

\section{Method}

\subsection{Preferred Velocity Correction}

So far, the calculation of preferred velocity is still based on a greedy rule, where each agent wishes to move in the direction that brings it closest to its destination in the shortest distance.
This holds true for an individual pedestrian, but in a crowd, individual preferences are often influenced by the collective \cite{le2002crowd}.
In real-life scenarios such as subway stations, airports, etc., the movement preferences of crowds often exhibit a tendency to follow the pedestrians in front, and even create channeling effects \cite{helbing1995social} in some scenes. These autonomously emerging crowd phenomena can often enhance the efficiency of crowd movement.
The basic idea in this work is to adjust the preferred velocity of each agent based on the surrounding crowd state, trying to create some beneficial crowd dynamics to enhance the efficiency of crowd movement.
Therefore, we modify the calculation of $\V{V}^t$ as follows:
\begin{equation}
\V{V}^{t+1} = \mathcal{F}\left( \V{X}^{t}, \V{V}^{t}, \V{G}^t \right).
\end{equation}
where $\V{G}^t$ is considered as the new preferred velocities of agents after being influenced by the surrounding crowd.
Each $\V{g}^t_i$ in $\V{G}^{t}$ for each agent $A_i$ should be a more promising direction for the agent when the surroundings are crowded, 
compared with its original preferred velocity $\V{p}_i^t$.
At the same time, $\V{g}^t_i$ must also possess the functionality of $\V{p}_i^t$, guiding the agent towards its (local) destination.

How to find a reasonable $\V{G}^t$ is the focus of our work.
The proposed method in this paper is to make a slight rotation, by an angle $\theta_i$, to the original preferred velocity of each agent $A_i$ based on its current neighboring agents.
Therefore, the relationship between $\V{G}^t$ and $\V{P}^t$ is as follows.
\begin{gather}
\V{G}^t = \V{R}^t \circ \V{P}^t ,\\
\V{R}^t = R\left(\Theta^t\right) = \left[ R\left(\theta_1^t\right),R\left(\theta_2^t\right), \cdots, R\left(\theta_i^t\right), \cdots, R\left(\theta_N^t\right)  \right],
\end{gather}
in which $R(\theta_i)$ donates the rotation matrix corresponding to the angle $\theta_i$, $\circ$ represents the Hadamard product \cite{horn2012matrix}.
In this case, the goal of this work is to calculate for $N$ offset angles $\Theta^t = [\theta_1^t, \theta_2^t, \cdots, \theta_N^t]$ at each time step $t$, which can be characterized by \eqref{eq:cal_thetas}.
\begin{equation}
\Theta^t = \mathcal{G}\left( \V{X}^{t}, \V{V}^{t}, \V{P}^t \right)
\label{eq:cal_thetas}
\end{equation}
In the next subsection, a simple and efficient method for producing $\Theta^t$ is presented.

\subsection{Factors Influencing the Rotation Angle}

Most existing crowd simulation models calculate $\V{v}_i^{t+1}$ for each agent $A_i$ based on several kinds of information, including the current velocity $\V{v}_i^t$ and position $\V{x}_i^t$ of $A_i$, the agent's preferred velocity $\V{p}_i^t$, and the positions and velocities of other agents around $A_j$.
While these pieces of information primarily serve the purpose of collision avoidance, they also have the potential to offer guidance for optimizing crowd movement efficiency—a facet that has not received adequate attention in prior research.
In this section, we examine the potentially viable offset direction for each agent, considering the information mentioned above.
By simulating common behavioral patterns observed in real-life crowds, we hope to construct a following effect among individuals, thereby achieving crowd implicit grouping.

\subsubsection{View field}

In real-world scenarios, the preferences of pedestrians are often influenced only by the states of the crowd in front of them (within their field of view).
Therefore, we introduce a term to model this phenomenon.
We define a vector $\Phi^t$ as the directions of the agents at the current time step $t$:
\begin{equation}
\Phi^t = \left[ \phi^t_1, \phi^t_2, \cdots, \phi^t_i, \cdots, \phi^t_N \right].
\end{equation}
In this work, we do not consider the actual velocity of an agent at time step t as its direction because the actual velocity may result in a very strange direction when avoiding collisions. 
Instead, we take a compromise between the actual velocity and the preferred velocity as the direction of an agent:
\begin{equation}
\Phi^t = \V{V}^t + \V{P}^t.
\end{equation}
Whether other agents are in front of agent $A_i$ can be inferred using the following formula:
\begin{equation}
T_0 = H\left( \left(\Phi^t\right)^\top \otimes \V{1}_{1\times N} \odot \Delta \V{X}^t \right),
\end{equation}
where $H(\cdot)$ is the Heaviside step function \cite{zhang2020feature}, $\Delta\V{X}^t$ is the matrix of relative positions of all agents:
\begin{equation}
\Delta \V{X}^t  = \V{X}^t \otimes \V{1}_{N\times 1} - \left(\V{X}^t\right)^\top \otimes \V{1}_{1\times N},
\end{equation}
in which $\otimes$ donates the Kronecker product \cite{henderson1983history}, $\odot$ donates element-wise dot product, $\V{1}_{n\times m}$ means an all-ones matrix with size of $n \times m$.
$T_0$ is a matrix of dimensions n-by-m, comprised of ones and zeros. 
The element in the $i$-th row and $j$-th column signifies whether $A_j$ is positioned in front of $A_i$, with one indicating yes and zero indicating no.
When calculating $\theta^t_i$, $T_0$ can be used as a mask to control which agents' information needs to be included in the calculation.

\subsubsection{The similarity and dissimilarity in directions}
The movement of a pedestrian is often influenced by the velocities of the surrounding crowd. 
The velocities of other pedestrians commonly serve as a foundation for achieving collision avoidance. 
Simultaneously, the preferred velocity of a pedestrian is also influenced by these velocities.
In real-world situations, when a pedestrian encounters a group of individuals walking towards them, there is a tendency to avoid them. Conversely, the pedestrian is inclined to follow a group of people moving in a similar direction as them.
Inspired by this phenomenon, this work uses this observation as partial basis to calculate $\Theta^i$.

The degree of difference in the direction between an agent $A_i$ and another agent $A_j$ can be calculated by their inner product $\V{\phi}^t_i \cdot \V{\phi}^t_j$.
In this work, we assume that agents can only know the actual velocities of other agents at time step t, without knowing their preferred velocities.
Therefore, for agent $A_i$, we use $\V{\phi}^t_i \cdot \V{v}^t_j$ as an indicator of the difference in velocity between $A_i$ and $A_j$.
When $\V{\phi}^t_i \cdot \V{v}^t_j$ is positive, the larger the value, the closer the directions of $A_i$ and $A_j$.
When $\V{\phi}^t_i \cdot \V{v}^t_j$ is negative, the larger the absolute value, the more conflicting the directions of $A_i$ and $A_j$.
These indicators form a $N \times N$ matrix, which can be calculated by \eqref{eq:cal_vv}.
\begin{equation}
T_1 = \left(\V{\Phi}^t\right)^\top \V{V}^t .
\label{eq:cal_vv}
\end{equation}

The value of $\V{\phi}^t_i \cdot \V{v}^t_j$ can be positive or negative. 
If it is negative, $A_i$ needs to avoid $A_j$; if it is positive, $A_i$ could follow $A_j$.
The magnitude of $\theta^t_i$, which represents the degree of correction to the preferred velocity of $A_i$, can partially depend on the absolute value of $\V{\phi}^t_i \cdot \V{v}^t_j$.

\subsubsection{Detour}

The magnitude of the angle between the direction of an agent $A_i$ and the relative position of another agent $A_j$ with respect to $A_i$ can serve as an indicator of the potential obstruction that $A_j$ might pose to the movement of $A_i$ in future time steps.
When there are numerous other agents in the front of an agent, it suggests the possibility of potential congestion ahead \cite{liao2023towards}. 
$A_i$ can make some detour-like adjustments, i.e., offsetting the preferred velocity.
In this work, this indicator is utilized as a component in calculating $\Theta^t$.
\begin{equation}
T_2 = \left(\Phi^t\right)^\top \otimes \V{1}_{1\times N} \odot \Delta \V{X}^t \oslash \sqrt{\Delta \V{X}^t \odot \Delta \V{X}^t},
\end{equation}
where $T_2$ is a $N \times N$ matrix, $\oslash$ donates Hadamard division \cite{cyganek2013object}.
The element in the $i$-th row and $j$-th column of $T_2$ signifies the degree of proximity between the movement direction of $A_i$ and the direction of $A_j$ relative to $A_i$.

\subsubsection{Distance}

A basic intuition is that distant agents have a reduced impact on oneself.
This work uses the reciprocal of the relative distance to describe this influence:
\begin{equation}
T_3 = \left(\sqrt{\Delta \V{X}^t \odot \Delta \V{X}^t}\right)^{\circ-1}.
\end{equation}
The impact on agent $A_i$ from other agents positioned at a greater distance should be diminished.
This idea has also been widely adopted in many previous studies, such as neighbor region \cite{van2008reciprocal}.
When adjusting the preferred velocity, it is also not necessary to consider all other agents.
A new mask $T_4$ can be used to achieve this:
\begin{equation}
T_4 = 1 - H\left( \sqrt{\Delta \V{X}^t \odot \Delta \V{X}^t} - L_{m} \right),
\end{equation}
where $L_{m}$ is the radius of the neighbor region. 
When the distance between $A_j$ and $A_i$ is greater than $L_{m}$, $A_j$ is not considered to have an impact on the preferred velocity of $A_i$.

\subsubsection{Rotation direction of preferred velocity}

The indicators introduced above can provide a basis for the magnitude of the preferred velocity offset angle, but we need to introduce a new component to control the direction of the preferred velocity offset.
This work proposes a method to control the rotation direction of the preferred velocity via cross product.
Because the cross product of $\V{x}^t_j-\V{x}^t_i$ and $\phi^t_i$, and the sine of the angle between them have a linear relationship, we can use their cross product to determine whether $A_j$ positioned in the left or right half area of $A_i$.
Therefore, we use $T_5$ to control the rotation direction of each preferred velocity:
\begin{equation}
T_5 = 2H\left( \left(\Phi^t\right)^\top \otimes \V{1}_{1\times N} \star \Delta \V{X}^t \right) - 1,
\end{equation}
where $\star$ donates element-wise cross product.
$T_5$ is a matrix consisting of $1$s and $-1$s. 
If the value in the $i$-th row and $j$-th column is $1$, it means that $A_j$ is on the left side of $A_i$; otherwise, on the right side.

\subsection{Overall Framework}
Based on the analysis in the previous subsection, we have obtained some useful information (i.e., $T_0$ to $T_5$) for calculating $\Theta^t$.
This subsection employs a straightforward multiplication approach to organize the potential information outlined earlier.
As $T_5$ is obtained via the cross-product operation that is associated with the sine function.
Its values are distributed around $0$ when the state of crowd is completely random.
Therefore, this work establishes the following relationship:
\begin{equation}
\sin\left(\Theta^t\right) \underset{\sim}{\varpropto} \V{1}_{1\times N} \overset{5}{\underset{i=0}\bigcirc}\left(T_i\right)^\top.
\end{equation}
We can finally obtain the calculation formula for $\Theta^t$ by normalization and the addition of the coefficient $K$:

\begin{equation}
\Theta^t = \sin^{-1}\left(\frac{1}{2}\tanh\left(K \V{1}_{1\times N} \overset{5}{\underset{i=0}\bigcirc}\left(T_i\right)^\top  \right)\right),
\end{equation}
where the term $\frac{1}{2}$ is used to control the deviation of the preferred velocity, ensuring it does not exceed $\frac{\pi}{2}$.

The overall framework of the algorithm is illustrated in Algorithm \ref{al_framework}.
The simulation initiates with the assignment of initial positions $\V{X}^0$ and velocities $\V{V}^0$ for the agents.
In typical situations, $\V{V}^0$ is composed of $N$ zero vectors, indicating that agents are initially stationary. 
The simulation proceeds through discrete time steps until reaching a predefined maximum time limit $T_{max}$. 
At each time step, the preferred positions $\V{P}^t$ for agents are calculated based on their global paths. 
Then, the rotation angles $\Theta^t$ of $\V{P}^t$ are calculated via the proposed method.
Accordingly, $\V{P}^t$ is rotated and the new preferred velocities $\V{G}^t$ of agents can be obtained.
The velocities of agents are updated by collision avoidance model $\mathcal{F}(\cdot)$ that considers the current positions $\V{X}^t$, velocities $\V{V}^t$, and rotated preferred velocities $\V{G}^t$. 
In this work, ORCA \cite{van2011reciprocal} is adopted as the foundational collision avoidance model. Given its emphasis on preferred velocity as the objective, it aligns seamlessly with our approach.
The positions are updated accordingly, and the current state of the simulation is visualized. 
This iterative process continues until the maximum time limit is reached. 

\begin{algorithm}[ht!]
    \caption{The Proposed Method}
    \label{al_framework}
    \BlankLine
    Initial $\V{X}^0$ and $\V{V}^0$\;
    
    $t \leftarrow 0$\;
    \While{$t < T_{max}$}{
        Calculate $\V{P}^t$ based on the global paths of agents\;
        $\Theta^t \leftarrow \sin^{-1}\left(\frac{1}{2}\tanh\left(K \V{1}_{1\times N} \overset{5}{\underset{i=0}\bigcirc}\left(T_i\right)^\top  \right)\right)$ \;
        $\V{G}^t \leftarrow R\left(\Theta^t\right) \circ \V{P}^t$ \;
        $\V{V}^{t+1} \leftarrow \mathcal{F}\left( \V{X}^{t}, \V{V}^{t}, \V{G}^t \right)$,\quad$\V{X}^{t+1} \leftarrow \V{X}^t + \V{V}^{t+1} \Delta t$ \;
        Visualize$(\V{X}^t)$, \quad $t \leftarrow t+1$ \;
    }
\end{algorithm}

\section{Implementation}
\subsection{Setting}
This work primarily focuses on the efficiency of crowd movement.
Therefore, the evaluation metrics in this paper include the dissimilarity between preferred velocity and actual velocity $M_1 = \frac{1}{2N}\Delta t \sum_{i}^{N} |\V{v}^t_i - \V{p}^t_i|^2$ and crowd congestion \cite{liao2023crowd} level
\begin{equation}
M_2 = \frac{1}{N^2} \sum_{i\neq j}^{N}\frac{1-\cos(\angle{\V{v}^t_i,\V{v}^t_j})}{2}e^{-|x^t_i-x^t_j|}.
\end{equation}

\begin{figure}
  \centering
  \begin{subfigure}{0.46\linewidth}
    \includegraphics[width=\linewidth]{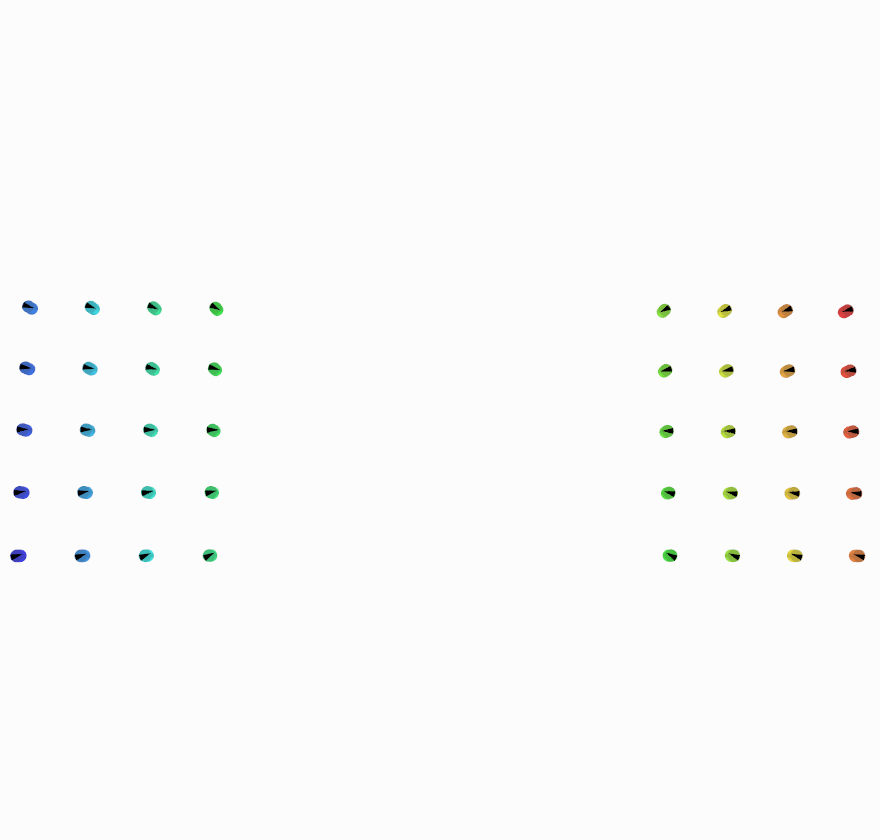}
    \subcaption{initial state of 2-Group}
    \label{fig:2_group}
  \end{subfigure}
  \hspace{0.1in}
  \begin{subfigure}{0.4\linewidth}
    \includegraphics[width=\linewidth]{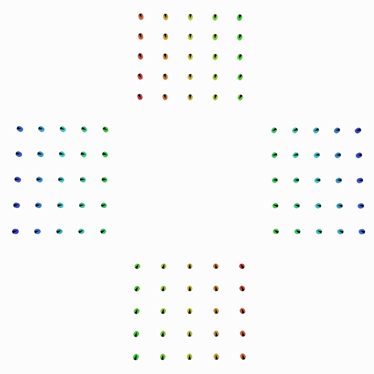}
    \subcaption{initial state of 4-Group}
    \label{fig:4_group}
  \end{subfigure}
  \caption{The initial distribution of agents of two scenarios}
\end{figure}

We test our method in several classic scenarios, including: 
\textbf{Circle} (see Figure \ref{fig:teaser}, a classic scenario involves the even distribution of 100 agents on a circle, with their goal being to navigate to the antipodal position on the same circle), 
\textbf{AsyCircle} (similar to the configuration of Circle scenario, but with a slight perturbation added to the destination of agents, causing their movement to be not entirely symmetrical),
\textbf{2-Group} (see Figure \ref{fig:2_group}, two groups, each comprising 20 agents, are positioned on the east and west sides of the scenario in a grid formation, with each agent's destination mirroring horizontally w.r.t. the scenario center), and
\textbf{4-Group} (see Figure \ref{fig:4_group}, four groups, each comprising 25 agents, are arranged in a grid formation in four directions of the scenario, and each agent's destination is a position symmetric w.r.t. the scenario center).

Our model is compared with following models: \textbf{SocialForces} \cite{helbing1995social}, \textbf{RVO} \cite{van2008reciprocal}, \textbf{PLEdestrians} \cite{guy2010pledestrians},  \textbf{PowerLaw} \cite{karamouzas2014universal}, \textbf{Karamouzas} \cite{karamouzas2010velocity}, \textbf{Moussaid} \cite{moussaid2011simple}, \textbf{ORCA} \cite{van2011reciprocal}, \textbf{Dutra} \cite{dutra2017gradient} and \textbf{Implicit} \cite{karamouzas2017implicit}.
We implemented our method in C++ for the simulation of agents and Python for user interface.
The parameters in our work are configured with $\Delta t = 0.1, r = 0.3, K = 0.6, L_m = 10$.

\subsection{Results and Analysis}

\begin{figure}[ht]
  \centering
  \includegraphics[width=\columnwidth]{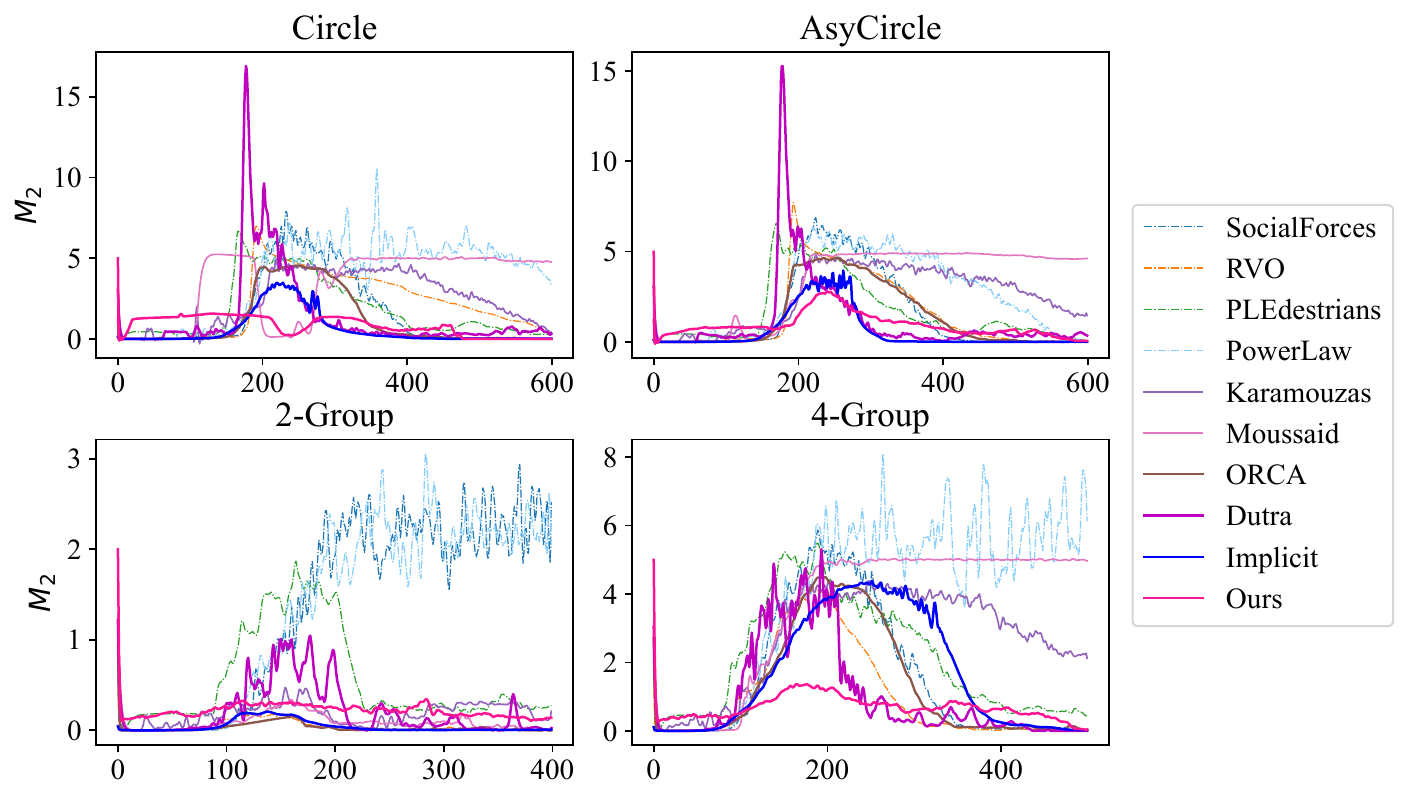}
  \caption{Curves of $M_1$ (the dissimilarity between preferred velocity and actual velocity of agents) over time steps of different crowd simulation models in four scenarios.
  In each scenario, although our approach exhibits a brief period of significant dissimilarity at the initial stage, it maintains relative stability throughout the subsequent simulation process.}
  \label{fig:m1}
\end{figure}

\begin{figure}[ht]
  \centering
  \includegraphics[width=\columnwidth]{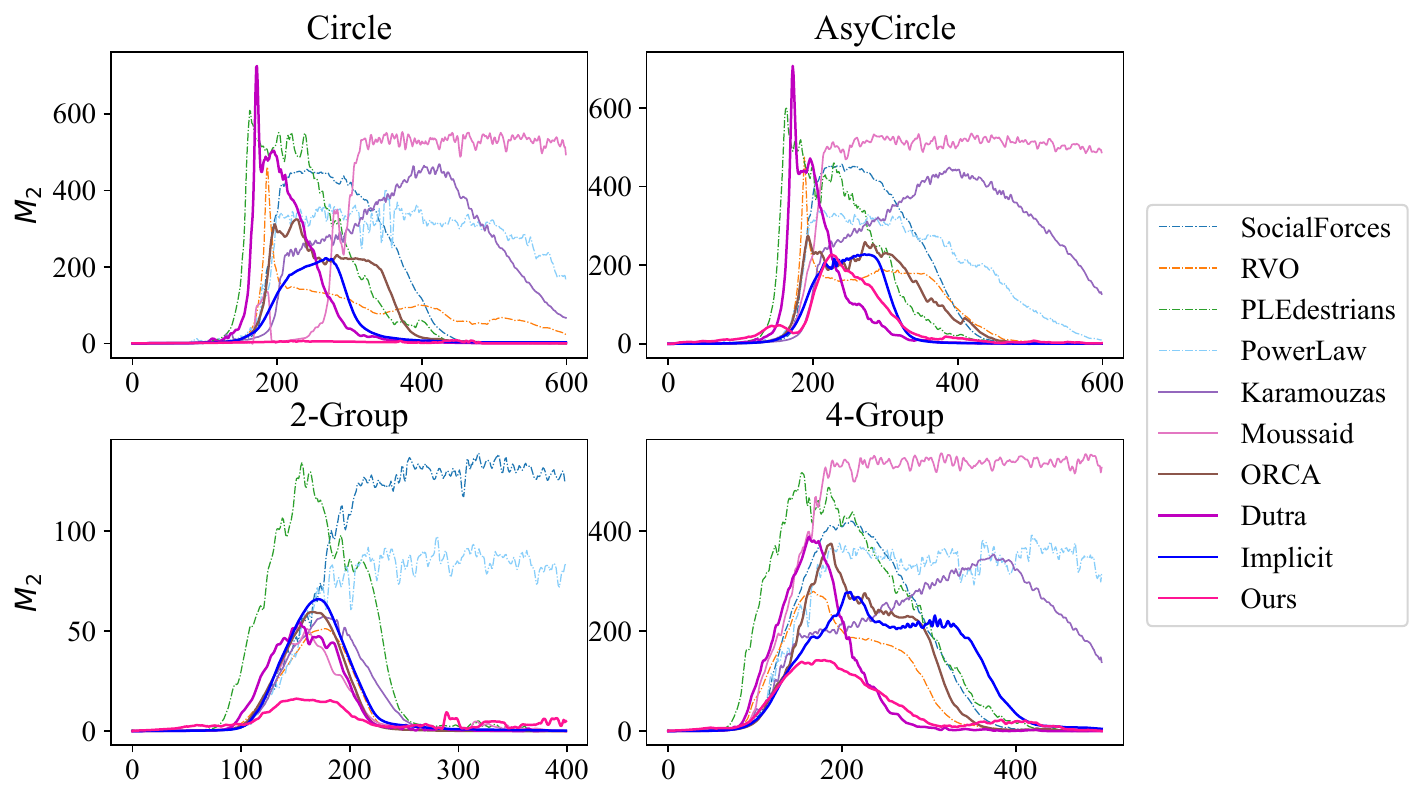}
  \caption{Curves of $M_2$ (the crowd congestion level) over time steps of different crowd simulation models in four scenarios. Our method consistently maintains a low level of congestion when agents encounter others (time steps around the middle of the x-axis)}
  \label{fig:m2}
\end{figure}

Figure \ref{fig:m1} presents the variation of $M_1$ over time steps in different scenarios.
It can be observed that when agents from different directions converge (time steps around the middle of the x-axis), the actual velocities of agents in most methods deviate significantly from their preferred velocities.
Our method achieved a relatively stable alignment between preferred velocity and actual velocity during the most of time of each simulation process.
This results indicate that applying beneficial rotations to the agents' original preferred velocities $\V{P}^t$ can actually reduce the dissimilarity between $\V{P}^t$ and $\V{V}^{t+1}$.
We hope that these results can provide some potential insights for future researchers in related studies.

Figure \ref{fig:m2} depicts the variation of congestion levels over time steps for different algorithms in different scenarios.
From the figure, it can be observed that our method exhibits relatively low congestion level when agents from different directions start to converge.
Surprisingly, in the Circle scenario, due to the highly symmetrical movement of the crowd, all agents self-organize into a single group (refer to Figure \ref{fig:teaser}), resulting in no congestion.
Implicit crowd also performed well in most scenarios.
However, the Dutra method performed poorly in two circle-based scenarios. 
Referring to the motion trajectories in Figure \ref{fig:teaser}, it can be observed that this vision-based method exhibits severe oscillations when encountering a large and densely packed number of agents from different directions.
SocialForces and PowerLaw are two rule-based methods that, without adding perturbations, exhibit prolonged congestion in symmetric scenarios.
The area enclosed by the curve and the x-axis can be approximated as the overall congestion level of the crowd throughout the simulation process.
Our method demonstrated the minimum congestion level in each scenario.
This indicates that mimicking the following effect of real-life crowd behavior can significantly reduce congestion among individuals.

\begin{figure}[ht]
  \centering
  \includegraphics[width=0.9\linewidth]{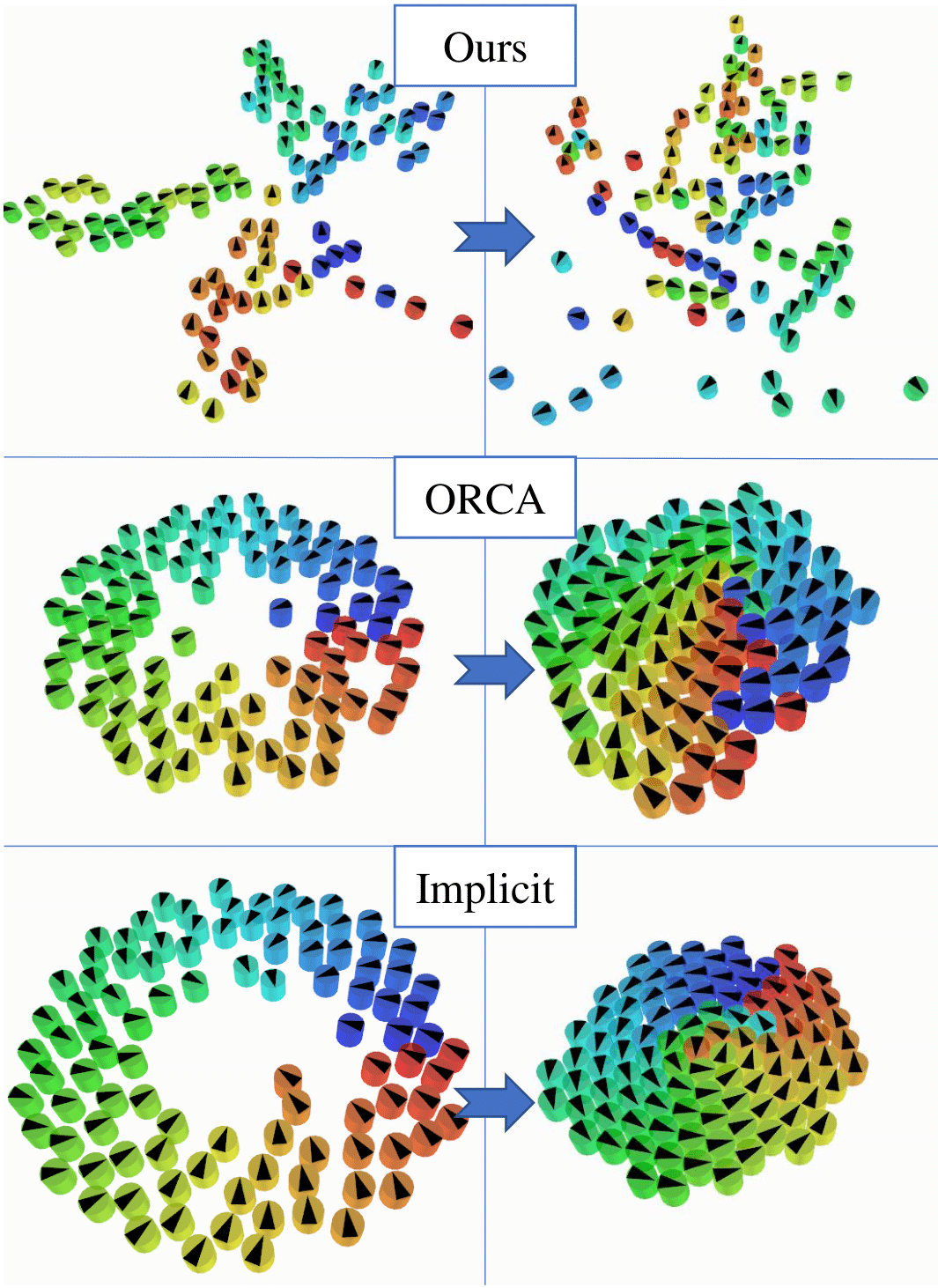}
  \caption{Agent states for three crowd simulation models in the AsyCircle scenario. 
  Both ORCA and Implicit exhibit congestion at the area where the agents converges in the middle.
  Our method results in loosely formed groups of agents with similar directions before they encounter each other. 
  Upon meeting, there is no significant congestion observed.}
  \label{fig:demo}
\end{figure}

\begin{figure}[ht]
  \centering
  \includegraphics[width=0.7\linewidth]{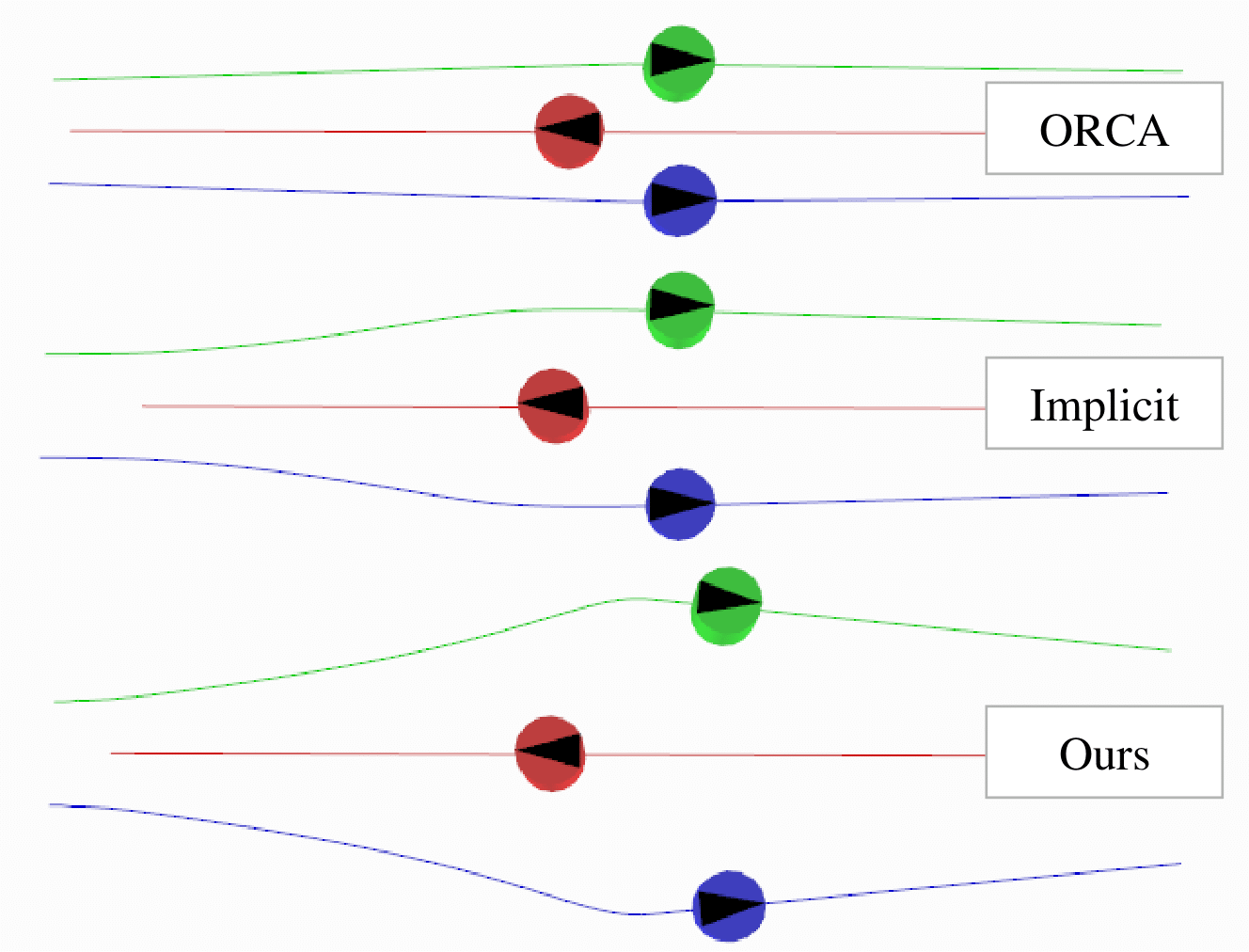}
  \caption{Agent states for three crowd simulation models in classic three-agent scenario. 
          Compared to the close encounters in ORCA, Implicit and our method demonstrate more realistic behaviors by incorporating necessary avoidance maneuvers and detours.}
  \label{fig:three_agents}
\end{figure}

We also compares three crowd simulation models – ORCA, Implicit, and our method – within the AsyCircle scenario visually (its agents' initial positions can be referenced from the first subplot in Figure \ref{fig:teaser}). 
Figure \ref{fig:demo} presents the simulation outcomes, highlighting distinctive characteristics of each model. 
Both ORCA and Implicit exhibit congestion at the convergence area where agents meet in the middle. 
In contrast, our proposed method showcases a unique behavior, forming loosely structured groups of agents with similar directions prior to their encounters. 
After agents meet, the original Implicit group disperses, and they reorganize into new formations with similar velocity directions (visible queues of agents in the figure).
Notably, in our crowd simulation method, there is a conspicuous absence of significant congestion, emphasizing the efficacy of our approach in mitigating crowd congestion issues.
Figure \ref{fig:three_agents} shows illustrates partial trajectories of three agents from different crowd simulation models in a classic three-agent scenario.
Agents in our method exhibit stronger proactive avoidance behavior compared to the Implicit and ORCA.
This indicates that the pattern in our method, which considers avoiding or taking detours around pedestrians ahead, works.

\begin{figure}[ht]
  \centering
  \includegraphics[width=\linewidth]{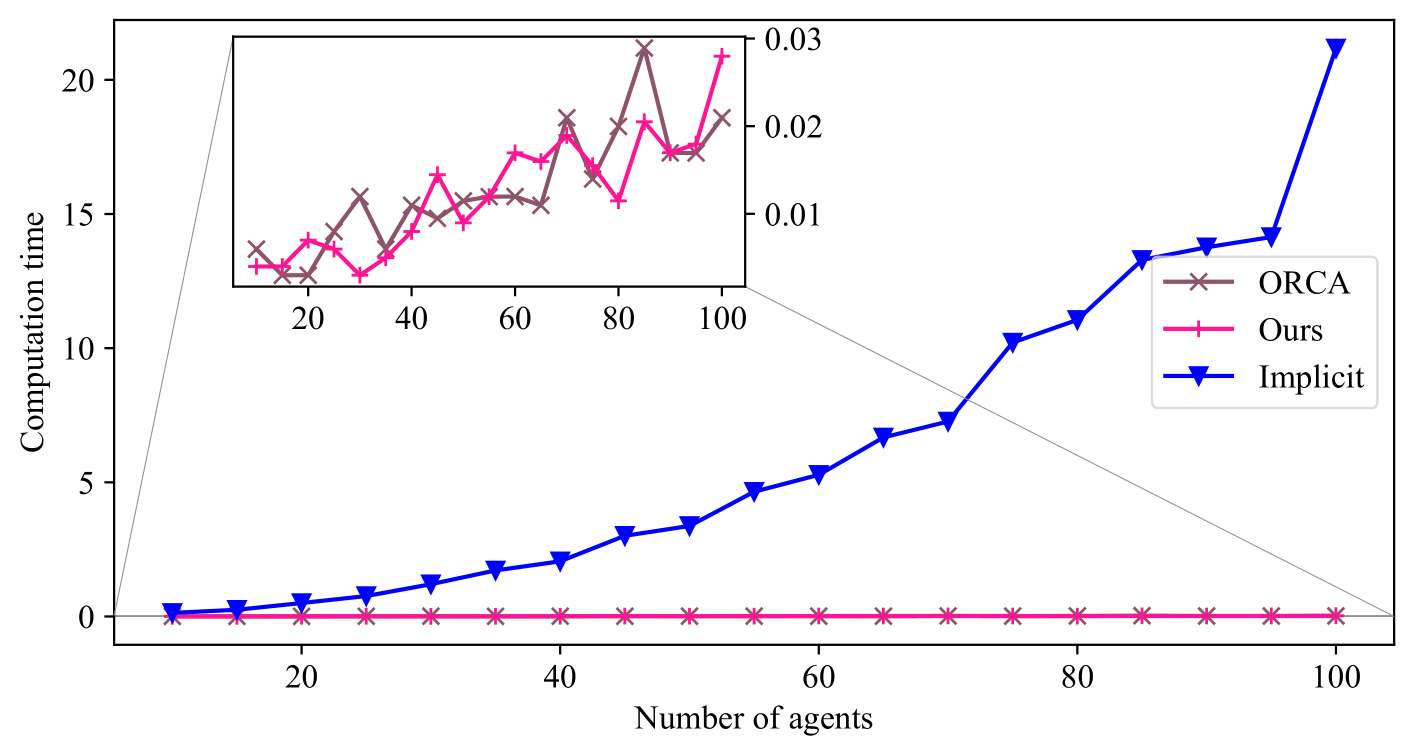}
  \caption{The computation time of the ORCA, Implicit, and our method for different numbers of agents. Our algorithm is highly efficient, introducing nearly negligible computational overhead to the underlying ORCA.}
  \label{fig:time_analy}
\end{figure}

Figure \ref{fig:time_analy} illustrates the computation time of the ORCA, Implicit, and our method for different numbers of agents.
From the figure, it can be observed that our algorithm is highly efficient, adding almost negligible computational overhead compared to the ORCA.
While Implicit is a global optimization algorithm, its runtime efficiency is comparatively lower.

\section{Conclusion}

In this paper, we propose an approach that allows crowd grouping to emerge via self-organization, leading to efficient crowd movement.
Experimental results demonstrate that our approach significantly reduces crowd congestion and the dissimilarity between agents' preferred and actual speeds, compared to several other crowd simulation models.
This indicates that our original intention to achieve efficient crowd movement via the autonomous emergence of latent behaviors of crowd, such as implicit group formation of agents with similar directions, is feasible.

Our method achieves the emergence of implicit groups, the process of loosely formed groups being disrupted and then reassembled (after unavoidable encounters with groups in severe directional conflict) is highly inefficient. Future research could focus on addressing this issue to achieve more efficient crowd movement.
While the components $T_0$ to $T_5$ of our method are built based on the rationale of real-world scenarios, the final combination of these components, however, is still at a very preliminary stage.
Moreover, the combination also introduces an additional hyperparameter, which is not an ideal design. 
In the future, considering the use of symbolic regression methods can potentially address the mentioned issues.

\bibliographystyle{ACM-Reference-Format}

\end{document}